\documentclass[aps,prd,reprint,a4paper,preprintnumbers,superscriptaddress,longbibliography,amsmath,amssymb,floatfix,nofootinbib]{revtex4-2}
\pdfoutput=1
\usepackage{graphicx}
\usepackage{xcolor,soul}
\usepackage{slashed}
\usepackage{enumitem}
\usepackage[colorlinks=true, linkcolor=blue, filecolor=blue, urlcolor=blue, citecolor=blue, pdftex, plainpages=false
]{hyperref}
\usepackage{orcidlink}
\newcommand{\msbar}{\overline{\mathrm{MS}}}
\newcommand{\MSbar}{$\overline{\mathrm{MS}}$}

\begin{document}
\title{Gradient Flow Renormalization Schemes for Composite Fermion Operators}

\date{\today}
\author {Matthew Black\orcidlink{0000-0002-8952-1755}}
\email{matthew.black@ed.ac.uk}
\affiliation{Higgs Centre for Theoretical Physics, School of Physics and Astronomy, University of Edinburgh, Edinburgh EH9 3JZ, UK}

\author{Anna Hasenfratz\orcidlink{0000-0003-1813-2645}}
\email{anna.hasenfratz@colorado.edu}
\affiliation{Department of Physics, University of Colorado Boulder, 
Boulder, Colorado 80309, USA}

\author{Oliver Witzel\orcidlink{0000-0003-2627-3763}}
\email{oliver.witzel@uni-siegen.de}
\affiliation{Theoretische Physik 1, Center for Particle Physics Siegen, Naturwissenschaftlich-Technische Fakult\"at, Universit\"at Siegen, 57068 Siegen, Germany}

\preprint{SI-HEP-2026-13, P3H-26-050}

\begin{abstract}
We introduce gradient flow (GF) normalization prescriptions for fermionic composite operators in which the flowed fermion wavefunction renormalization factor is fixed nonperturbatively  using either the partially conserved axial charge or the conserved vector current. The resulting $A$ and $V$ schemes are defined through standard flowed two-point correlation functions and therefore avoid the backward-flow construction required by local ringed-scheme definitions. In the short-flow-time limit, the $A$ and $V$ schemes can be matched to $\overline{\mathrm{MS}}$ using known ringed-scheme short-flow-time expansion (SFTX) coefficients. 
We show how these schemes can be implemented through ratios of two-point correlation functions, leading to simple nonperturbative determinations of renormalization factors, anomalous dimensions, and evolution factors
which connect lattice-accessible flow times to shorter flow times where perturbative matching is reliable.
We illustrate the method with RBC-UKQCD domain-wall fermion ensembles, including a GF determination of the ratio of matching factors $Z_V/Z_A$, and a new GF determination of the renormalized strange quark mass.
\end{abstract}
\maketitle

\section{Introduction}

The gradient flow (GF) has emerged as a powerful tool for defining renormalized observables in quantum field theory \cite{Luscher:2010iy,Luscher:2013cpa}.
By evolving the fields along the flow time $\tau$, ultraviolet fluctuations are suppressed, and the fields are smoothed over a radius $\sqrt{8\tau}$. 
Flowed gauge observables are finite and renormalized  at an energy scale $\mu\propto 1/\sqrt{8\tau}$. Fermions require   multiplicative wavefunction renormalization ${\cal Z}_\chi(\tau)$ but composite operators built from flowed fields do not require additional renormalization.
Composite operators containing $n$ fermion fields renormalize according to
\begin{align}
O_{\mathrm{GF}}(\tau;x)
= {\cal Z}_\chi^{n/2}(\tau)\, O(\tau;x),
\end{align}
where $O(\tau,x)$ is constructed from flowed fields. Matching to continuum schemes such as \MSbar{} is possible through the short-flow-time expansion (SFTX)  \cite{Luscher:2011bx,Suzuki:2013gza,Borgulat:2023xml},
\begin{align}
O_{\overline{\mathrm{MS}}}(\mu;x)
= \lim_{\tau\to 0} \zeta_O^{-1}(\mu,\tau)\, O_{\mathrm{GF}}(\tau;x),
\end{align}
which provides a perturbative connection between flowed and continuum operators. 
The scheme dependence of GF renormalization is entirely encoded in the definition of ${\cal Z}_\chi(\tau)$. 

Gradient flow renormalization has been successfully applied in a wide range of contexts, including the determination of the renormalization group (RG) $\beta$ functions~\cite{Fodor:2017die,Hasenfratz:2019hpg}, running couplings and $\Lambda$ parameters~\cite{DallaBrida:2018rfy,Bruno:2017gxd,Hasenfratz:2023bok,Wong:2023jvr,DallaBrida:2026kuo}, quark masses~\cite{Black:2025gft}, the energy-momentum tensor~\cite{Suzuki:2013gza,Makino:2014taa}, 
bag parameters for neutral meson mixing or lifetimes of heavy mesons~\cite{Suzuki:2020zue,Black:2026rbz,Black:2026dzp}, parton distribution functions (PDFs)~\cite{Francis:2025rya,Francis:2025pgf,Edwards:2026ixr},  and related quantities. 
In many of these applications, ratios of correlation functions are constructed such that the wavefunction renormalization ${\cal Z}_\chi(\tau)$ cancels. 
While this approach has proven effective, it also limits the class of observables that can be treated. In particular, the explicit determination of ${\cal Z}_\chi(\tau)$ is required whenever ratios cannot be formed, or absolute normalization is needed, such as in the case of the nonperturbative determination of anomalous dimensions.

A well-defined prescription for ${\cal Z}_\chi(\tau)$ is provided by Suzuki's  ``ringed'' scheme, in which the renormalization factor $\mathring{\cal Z}_\chi(\tau)$ is defined through expectation values of  flowed local operators  \cite{Suzuki:2013gza}. 
While this scheme is convenient in perturbation theory and has a well-understood matching to \MSbar{}, its direct implementation in lattice simulations is computationally demanding, requiring adjoint or ``backward'' GF \cite{Luscher:2013cpa}.

In this work, we formulate alternative GF renormalization schemes that are tailored to lattice calculations. 
Instead of relying on the expectation values of local composite operators, we introduce renormalization schemes in which the normalization of the partially-conserved axial charge or conserved vector current is prescribed to remain invariant under the evolution in flow time. 
We refer to these prescriptions as $A$ and $V$ schemes. 
The corresponding wavefunction renormalization factors $\widetilde{\cal Z}_\chi^{(A,V)}(\tau)$ can be determined from simple two-point correlation functions, making them numerically efficient and straightforward to implement.

The axial, vector, and ringed schemes are mutually equivalent in the short-flow-time limit, and thus the SFTX coefficients obtained in the ringed scheme can be reused to convert the axial and vector schemes to \MSbar{}.

A central challenge in GF–based renormalization is the simultaneous presence of several competing physical scales. The lattice spacing $a$ and the quark mass $m_q$ constrain the admissible flow-time window:
\begin{align}
\sqrt{8\tau} \gg a,
\end{align}
is necessary to control flow-time cutoff effects, while 
\begin{align}
\sqrt{8\tau} \, \ll \frac{1}{m_q},
\end{align}
ensures that all infrared scales remain parametrically small compared to the flow time scale $\mu \propto 1/\sqrt{8\tau}$.

The GF $A$ and $V$ schemes enable the extraction of scale-dependent anomalous dimensions within the GF framework. 
This, in turn, provides a fully nonperturbative procedure to relate operator expectation values defined at the lattice  scale to higher energy scales where perturbative matching to continuum schemes is more robust. 
The RG running used here thus offers a systematic approach to bridging the separation between these disparate scales.

This paper is organized as follows. 
In Sec.~\ref{sec:GF} we review the gradient flow and introduce the fermion renormalization schemes.
The nonperturbative determination of the anomalous dimension and the RG evolution via GF are discussed in Sec.~\ref{sec:anom_dim}. In Sec.~\ref{Sec.Applications}, we  present three applications: 
prediction of the ratio of matching factors $Z_V/Z_A$; 
the nonperturbative determination of the anomalous dimension $\gamma_m$, needed for RG evolution; 
an alternative determination of the renormalized strange quark mass $m_s^{\msbar}$, where \MSbar{} matching is improved using nonperturbative flow time evolution via $\gamma_m$.
We conclude in Sec.~\ref{Sec.Conclusion}. Further details on the used gauge field ensembles are gathered in Appendix \ref{Sec.Setup}, whereas Appendix \ref{Sec.ZeroFlowTime} provides details on our procedure for extrapolating flowed results to the zero-flow-time limit in the SFTX.

\section{Gradient flow renormalization}
\label{sec:GF}

In this section, we briefly review GF for gauge and
fermion fields and introduce the renormalization of composite fermion
operators. 
We then define practical nonperturbative renormalization
schemes based on the vector and axial currents.

\subsection{Gradient flow for gauge and fermion fields}

GF defines a smoothing transformation of the fundamental
fields over the flow time $\tau$. The flowed gauge field
$B_\mu(\tau;x)$ is defined by
\begin{align}
  \begin{aligned}
    \partial_\tau B_\mu(\tau;x) &= D_\nu G_{\nu\mu}(\tau;x),\\
    B_\mu(0;x)&=A_\mu(x),
  \end{aligned}
\end{align}
where $G_{\mu\nu}$ and the covariant derivative $D_\mu$ are constructed from $B_\mu$ and the regular gauge field $A_\mu(x)$ serves as the initial condition at flow time $\tau=0$. 
The flow suppresses ultraviolet modes with momenta $p^2 \gg 1/\tau$ and smooths the gauge field over a radius $\sqrt{8\tau}$.

The flowed fermion fields $\chi(\tau,x)$ and $\bar\chi(\tau,x)$ are
defined by
\begin{align}
  \begin{aligned}
    \partial_\tau \chi(\tau;x)&=\Delta\,\chi(\tau;x),\\
    \partial_\tau \bar\chi(\tau;x)&=\bar\chi(\tau;x)\,\overleftarrow{\Delta},
  \end{aligned}
  \label{eq:fermionflow}
\end{align}
with initial conditions
\begin{align}
  \chi(0;x)=\psi(x),  \qquad
  \bar\chi(0;x)=\bar\psi(x),
\end{align}
where $\Delta=D_\mu D_\mu$ is the covariant Laplacian built from the flowed gauge field.

Unlike flowed gauge fields, flowed fermion fields need a
multiplicative wavefunction renormalization 
\begin{align}
  \begin{aligned}
    \chi_R(\tau;x)={\cal Z}_\chi^{1/2}(\tau)\,\chi(\tau;x),\\
    \bar\chi_R(\tau;x)={\cal Z}_\chi^{1/2}(\tau)\,\bar\chi(\tau;x)\, .
  \end{aligned}
  \label{eq:quarkren}
\end{align}
However, no additional renormalization is required for composite fermion operators.

\subsection{GF renormalization of fermion operators}
\label{sec:GF_scheme}

Consider a local fermion bilinear operator
\begin{align}
  O(x)=\bar\psi(x)\Gamma\psi(x),
\end{align}
and the corresponding flowed operator 
\begin{align}
  O(\tau;x)=\bar\chi(\tau;x)\Gamma\chi(\tau;x).
\end{align}  
We define the GF-renormalized operator as 
\begin{align}
  O_{\rm GF}(\tau;x)
  = {\cal Z}_\chi(\tau)\,O(\tau;x).
\label{eq:GFdef}
\end{align}
The wavefunction renormalization factor ${\cal Z}_\chi(\tau)$ can be defined in terms of expectation values of lattice operators. Different choices of ${\cal Z}_\chi(\tau)$ define different GF schemes.

\subsubsection{The ringed scheme}

A well-defined prescription is provided by Suzuki's ``ringed''
scheme \cite{Makino:2014taa}. 
In this scheme $\mathring{\cal Z}_\chi(\tau)$ is defined via the expectation value of the local operator $\langle \bar\chi(\tau,x)\,\overleftrightarrow{\slashed D}\,\chi(\tau,x)\rangle$, reading
\begin{align}
  \mathring{\cal Z}_\chi(\tau)
  = \sqrt{ \frac{-2\,\dim(R)\,N_f}{(4\pi\tau)^2\; \langle \bar\chi(\tau;x)\,\overleftrightarrow{\slashed D}\,\chi(\tau;x)\rangle}} \, ,
\label{eq:ringedZchi}
\end{align}
where $\dim(R)$ is the dimension of the fermion representation.

The ringed scheme can be matched to the \MSbar{} scheme
using perturbatively-calculable matching coefficients in the SFTX \cite{Borgulat:2023xml},
\begin{align}
  O_{\overline{\mathrm{MS}}}(\mu)
  &= \lim_{\tau \to 0} \mathring\zeta_O^{-1}(\mu,\tau)\,  \mathring O_{\mathrm{GF}}(\tau) \notag\\
  &= \lim_{\tau \to 0} \mathring\zeta_O^{-1}(\mu,\tau)\,  \mathring{\cal Z}_\chi(\tau)\,O(\tau)\,.
  \label{eq:ringed_matching}
\end{align}

The  ringed scheme is convenient because $\mathring{\cal Z}_\chi(\tau)$ is defined  in terms of local flowed fields, so the scheme admits a direct continuum perturbative treatment and short-flow-time matching to \MSbar{}.
However, the nonperturbative lattice
determination of $\mathring{\cal Z}_\chi(\tau)$ requires evaluating a local flowed operator,
which involves backward flow and is numerically demanding
\cite{Luscher:2013cpa}.

\subsubsection{The GF axial and vector scheme}

An alternative definition can be obtained by fixing the normalization of the axial or vector currents. 
The basic idea is simple:  the anomalous dimension of the conserved current vanishes, and so its flow-time dependence is entirely due to the wavefunction renormalization. Therefore, the conserved currents can be used to define the wavefunction renormalization factors \cite{Carosso:2018bmz, Hasenfratz:2022wll,Hasenfratz:2023sqa,Hasenfratz:2023wbr}. 

It is possible to define renormalization schemes where the axial charge $A_0 = \bar\psi \gamma_0 \gamma_5 \psi$ or the vector current $V_i = \bar \psi \gamma_i \psi$ does not change with the renormalization scale.
The \MSbar{} scheme is an example of such a renormalization scheme.  
In GF renormalization, we cannot impose this condition directly; rather, we require that the GF-renormalized axial charge or vector current, projected onto the appropriate lowest-lying physical state, is independent of the flow time $\tau$. 

To simplify the notation, we define zero-momentum mixed flowed-unflowed correlators.
\begin{align}
  C_{OO'}(\tau;t)
  = \sum_{\vec x}  \left\langle O(\tau;\vec x,t)\, O'(0;\vec 0,0) \right\rangle ,
  \label{eq:COtau}
\end{align}
and introduce the large-distance ratio
\begin{align}
  R_{O}(\tau)
  = \lim_{t\to\infty} \frac{C_{OO'}(0;t)}{C_{OO'}(\tau;t)} \, ,
  \label{eq:ROdef}
\end{align}
where $O'$ is an arbitrary probe that couples to the operator $O$, and $t$ denotes the Euclidean time.  
In the $t\to \infty$ limit of Eq.~\eqref{eq:ROdef} both correlators project onto the same physical state, and therefore the resulting ratio no longer depends on the Euclidean time $t$. 
The contribution of the probe $O'$ also cancels in the ratio.
In short, $R_O(\tau)$ measures the change in the overlap of the flowed operator $O(\tau;x)$ with the ground state relative to the unflowed operator.

We define the wavefunction renormalization factor in the GF axial ($A$) scheme as
\begin{align}
  \widetilde{\cal Z}_\chi^{(A)}(\tau)
  = Z_A\,R_{A}(\tau),
  \label{eq:ZchiA}
\end{align}
or, equivalently,
\begin{align}
  \widetilde{\cal Z}_\chi^{(A)}(\tau) \,\left\langle A_0(\tau)\, A_0(0) \right\rangle = 
    Z_A \, \left\langle A_0(\tau=0)\, A_0(0) \right\rangle \, ,
    \label{eq:ZchiA2}
 \end{align}
where all operators are taken at zero spatial momentum, and the  $t \to \infty$ limit is implicit on both sides of the equation. $Z_A$ is the  matching factor between the lattice axial charge operator and the partially-conserved axial charge predicted by Ward identities,
\begin{align}
  A_0^\text{cons} = Z_A A_0^\text{lat}\,.
\end{align}
$Z_A$ can be obtained by standard techniques and does not depend on the GF scheme. 
\begin{figure}[tb]
  \includegraphics[width=0.48\textwidth]{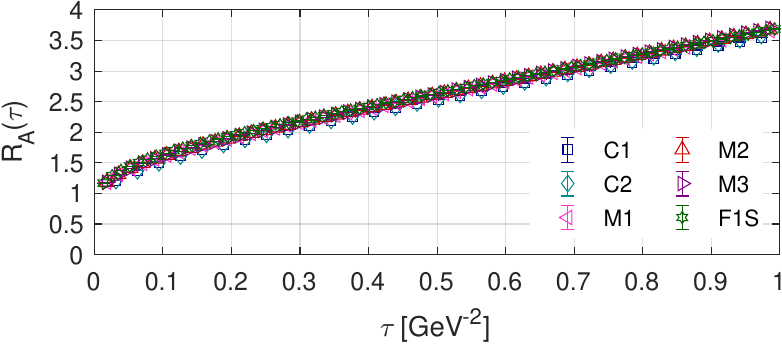}\\
  \includegraphics[width=0.48\textwidth]{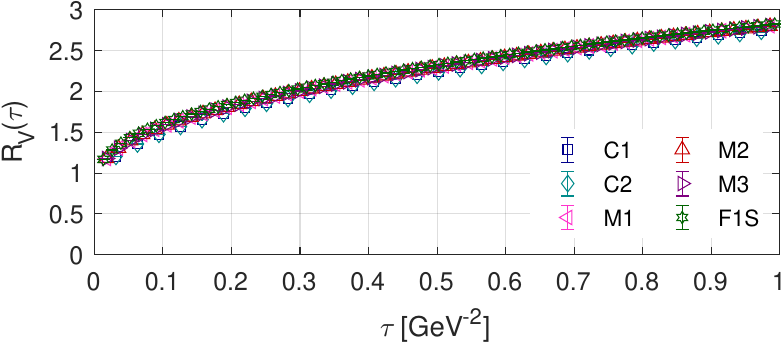}
  \caption{$R_A(\tau)$ (top) and $R_V(\tau)$ (bottom) determined using $(s\bar s)$-correlators vs.~the physical flow time $\tau$ $[\text{GeV}^{-2}]$. The ratios $R_A(\tau)$ and $R_V(\tau)$ are related to the GF-scheme renormalization factors as $R_A(\tau) = Z_A^{-1}\widetilde Z^{(A)}_\chi(\tau)$ and $R_V(\tau) = Z_V^{-1}\widetilde Z^{(V)}_\chi(\tau)$.}
  \label{Fig.Zchi_vs_tau_GeV2}
\end{figure}
Equation \eqref{eq:ZchiA} defines the \(A\)-scheme where, as intended, the GF-renormalized axial charge, projected onto the appropriate lowest-lying physical state, is independent of the flow time $\tau$.

Analogously using the conserved vector current, we can define the vector ($V$) GF scheme. 
In numerical simulations, the $V$ scheme has the advantage that the correlators are finite even in the chiral limit --- a property that is important if simulations with vanishing mass are required, such as when determining the anomalous dimension from small-volume simulations in the deconfined phase.
The $A$ scheme is advantageous in confined, finite-mass simulations since $A_0$ couples to the pseudoscalar state; therefore, its correlation function can be determined with high precision. 
The $A$ scheme also avoids the possible complication at lighter masses when the vector decays into two pions.\footnote{A scheme similar to the $V$ scheme is suggested in Ref.~\cite{Borgulat:2026onj}. The {\it caron} scheme is defined through the expectation value of the vector operator. One realization is suggested in Ref.~\cite{Shindler-talk} where the wavefunction renormalization factor is defined via the three-point function $\langle \pi | V_\mu | \pi \rangle$.} 

In this work, the Ward-identity normalization factors $Z_A$ and $Z_V$ are understood to be evaluated for the same valence action and quark mass as the correlators entering $R_O(\tau)$.
The resulting GF $A$ and $V$ schemes are therefore mass-dependent schemes at finite flow time. This differs from mass-independent schemes such as \MSbar{} or RI/MOM \cite{Martinelli:1994ty} in the chiral limit, but is natural for the hadronic correlator normalization used here.

In Fig.~\ref{Fig.Zchi_vs_tau_GeV2}, we show the ratios $R_A(\tau)$ and $R_V(\tau)$, which are respectively proportional to $\widetilde {\cal Z}_\chi^{(A)}(\tau)$ and $\widetilde{\cal Z}_\chi^{(V)}(\tau)$ as defined in Eq.~\eqref{eq:ROdef} for the $A_0$ and $V$ operators. 
These ratios are determined using a set of domain-wall fermion ensembles generated by the RBC-UKQCD collaboration, and the GF correlators~\cite{Black:2026data} which were generated for the projects of Refs.~\cite{Black:2025gft,Black:2026dzp, Black:2026rbz}. The properties of the ensembles are summarized in Appendix~\ref{Sec.Setup}, Table~\ref{Tab.Ensembles}.
Note that $\widetilde {\cal Z}_\chi^{(A,V)}(\tau)$ is divergent and does not have an $a\to 0$ continuum limit. 
The dependence on $a$ is logarithmic, and Fig.~\ref{Fig.Zchi_vs_tau_GeV2} shows only a small variation between the ensembles.

\subsection{Matching to \texorpdfstring{\MSbar{}}{MSbar}}
Operators renormalized in the GF $A$-scheme
\begin{align}
  \widetilde{O}^{(A)}_{GF}(\tau) = \widetilde{\cal Z}^{(A)}_\chi(\tau)\, O(\tau)\,,
  \label{eq:OA}   
\end{align}
can be matched to \MSbar{} in the short-flow-time limit as
\begin{align}
  O_{\msbar}(\mu) = \lim_{\tau\to 0}\widetilde\zeta^{-1}_O(\mu,\tau)\, \widetilde{O}^{(A)}_{GF}(\tau) \,,
  \label{eq:ZA_matching}   
\end{align}
where $\widetilde\zeta_O(\mu,\tau)$ denotes the corresponding SFTX matching coefficient that is predicted in the massless short-flow-time expansion. At finite valence quark mass, the same leading coefficient applies provided $\tau m_q^2 \ll 1$; the remaining
mass dependence is contained in higher-dimensional terms, beginning at ${\cal O}(\tau m_q^2)$, and is removed by the $\tau\to0$ limit.

In schemes such as the \MSbar{}, the conserved nonsinglet axial current has vanishing anomalous dimension, so its normalization is independent of the renormalization scale, and we can identify.
\begin{align}
  A_{0}^{\msbar}(\mu) = A_{0}^{\text{cons}} \, .
\end{align}
The GF $A$-scheme  is defined by imposing the analogous normalization on the large-distance axial-charge correlator: after multiplication by $\widetilde{\cal Z}^{(A)}_\chi(\tau)$, the flowed axial charge has the same projected matrix element as $A_0^{\rm cons}$ (see  Eqs.~\eqref{eq:ZchiA} and~\eqref{eq:ZchiA2}).
Consequently, the SFTX coefficient connecting the axial charge in the GF $A$ scheme to the conserved axial charge is trivial, i.e.
\begin{align}
  \widetilde\zeta_A(\mu,\tau)=1 + \mathcal{O}(\tau) \,.
\end{align}
Combining Eqs.~\eqref{eq:ringed_matching} and \eqref{eq:ZA_matching} for $O=A_0$ yields\footnote{Here $\mathring{\cal Z}_\chi$ and $\widetilde{\cal Z}^{(A)}_\chi$ are GF renormalization factors, while the explicit $\mu$-dependence resides in the perturbative matching coefficient $\mathring\zeta_A(\mu,\tau)$; any residual $\mu$-dependence at finite order is part of the perturbative matching uncertainty.}
\begin{align}
  \mathring{\cal Z}_\chi(\tau) = \mathring{\zeta}_A(\mu,\tau)\, \widetilde{\cal Z}_\chi^{(A)}(\tau) + \mathcal{O}(\tau, m_q^2)\,.
\end{align}
and for a general bilinear operator
\begin{align}
  \widetilde{\zeta}_O(\mu,\tau) = \mathring{\zeta}_O(\mu,\tau)\, \mathring{\zeta}^{-1}_A(\mu,\tau) + \mathcal{O}(\tau)\, .   
\end{align}
Substituting this into Eq.~\eqref{eq:ZA_matching}, we obtain
\begin{align}
  O_{\msbar}(\mu)
  &=\lim_{\tau\to 0} \, \mathring\zeta^{-1}_O(\mu,\tau)\,\mathring\zeta_A(\mu,\tau) \,\widetilde{\cal Z}^{(A)}_\chi(\tau)\, O(\tau)  \,,
  \label{eq:ZA_matching1}
\end{align}
i.e.~the ringed-scheme matching coefficients can also be used to connect the GF $A$ scheme to the \MSbar{} scheme.

The $V$-scheme is matched to \MSbar{} in the same way, resulting in
\begin{align}
  O_{\msbar}(\mu)
  &=\lim_{\tau\to 0}\, \mathring\zeta^{-1}_O(\mu,\tau)\,\mathring\zeta_V(\mu,\tau) \,\widetilde{\cal Z}^{(V)}_\chi(\tau)\, O(\tau)  \,.
  \label{eq:ZV_matching}
\end{align}

The GF $A$ and $V$ schemes thus provide simple, fully nonperturbative definitions of the wavefunction renormalization factors $\widetilde{\cal Z}^{(A,V)}_\chi(\tau)$ in terms of two-point correlation
functions that are numerically straightforward to evaluate. 
Both schemes can be matched to the \MSbar{} scheme using existing SFTX ringed-scheme coefficients. 
The above definitions can be applied in any context where a lattice operator must be renormalized and matched to the \MSbar{} scheme.

\section{Anomalous dimension and RG evolution}
\label{sec:anom_dim}

Gradient flow renormalization  provides a natural framework for
determining the anomalous dimension of composite operators directly from lattice correlation functions. 

Renormalization factors for lattice operators in GF schemes can be defined via expectation values. For an operator $O$, we define ${\cal Z}_O^{(A)}(\tau)$ as its renormalization factor in the $A$-scheme via
\begin{align}
  \langle O_\mathrm{GF}^{(A)}(\tau) O' \rangle = \widetilde{\cal Z}_O^{(A)}(\tau) 
  \langle O(\tau=0) O' \rangle,
  \label{eq:ZOdef}
\end{align}
where, as before, all operators are taken at zero spatial momentum, and the  $t \to \infty$ limit is implicit on both sides of the equation. Combining Eqs.~\eqref{eq:ZchiA}, \eqref{eq:OA} and \eqref{eq:ZOdef} yields
\begin{align}
  \widetilde{\cal Z}_O^{(A)}(\tau) 
  = \widetilde{\cal Z}_\chi^{(A)}(\tau) 
  \frac{1}{R_O^{(A)}(\tau)} 
  = Z_A \frac{R_A(\tau)}{R_O(\tau)} \,.
\end{align}
where $Z_A$ is the Ward-identity normalization factor at the quark mass value entering the correlator ratios.
Thus the renormalization factor $\widetilde{\cal Z}_O^{(A)}(\tau)$ describes the flow-time evolution of the mass-dependent renormalized operator $\widetilde{O}_\mathrm{GF}^{(A)}(\tau)$. 
The logarithmic flow-time dependence of $\widetilde{\cal Z}_O^{(A)}$ can be used to define the anomalous dimension of the operator $O$ as a function of $\tau$:
\begin{align}
  \widetilde \gamma^{(A)}_{O}(\tau)
  = -\,2\tau\frac{d}{d\tau} \log \widetilde{\cal Z}^{(A)}_{O}(\tau)
  = -\,2\tau\frac{d}{d\tau} \log\frac{R_A(\tau)}{R_O(\tau)}\,,
  \label{eq:gamma_tau}
\end{align}
which follows the standard field theory definition of the anomalous dimension
\begin{equation}\label{eq:flowed_gamma_def}
    \gamma\,G\,= {\mu}\frac{d\,G}{d\,\mu} = -\,2{\tau}\frac{d\,G}{d\,\tau},
\end{equation}
with the relation $\mu\frac{d}{d\mu} = -2\tau\frac{d}{d\tau}$.
At finite valence quark mass, $\widetilde \gamma^{(A)}_{O}(\tau)$ depends on the mass, but
this dependence is suppressed in the short-flow-time regime by powers of $\tau m_q^2$. The mass-independent RG anomalous dimension is obtained by taking the chiral limit.
In the short-flow-time limit, matching both the ringed and $A$-schemes to the same \MSbar{} operator implies
\begin{align}
  \widetilde \gamma^{(A)}_{O}(\tau)
  &= -\,2\tau\frac{d}{d\tau} \log \frac{\mathring{\zeta}_O(\mu,\tau)}{\mathring{\zeta}_A(\mu,\tau)} + {\cal O}(\tau)\, \notag \\
  &= \mathring{\gamma}_O(\tau) - \mathring{\gamma}_A(\tau)  + \mathcal O(\tau)\, ,
  \label{eq:ringed_relation}
\end{align}
where $\mathring\gamma_O\equiv-2\tau\frac{d}{d\tau}\log\mathring{\zeta}_O$ denotes the anomalous dimension in the ringed scheme~\cite{Harlander:2020duo,Borgulat:2023xml}.\footnote{Please note in Refs.~\cite{Harlander:2020duo,Borgulat:2023xml} the anomalous dimensions $\widetilde \gamma_O$ are defined in the ringed scheme directly as the logarithmic derivatives with respect to flow time, which differs by a factor of $-2$ from the definition in Eq.~\eqref{eq:flowed_gamma_def}. The use of $\mathring{\gamma}_O$ in this work requires multiplying the results of Refs.~\cite{Harlander:2020duo,Borgulat:2023xml} by $-2$.}

Equation \eqref{eq:gamma_tau} defines a  nonperturbative,
scale-dependent anomalous dimension that describes the RG evolution of the matrix element of the operator $\widetilde O^{(A)}_\mathrm{GF}$ between two flow times $\tau^* < \tau$, i.e.
\begin{align}
  \langle \widetilde O_{GF}^{(A)}(\tau^\ast)\, O(0) \rangle =
  C_m(\tau^\ast,\tau) \langle \widetilde O_{GF}^{(A)}(\tau)\, O(0) \rangle
\, ,
\label{eq:Otau}
\end{align}
where the flow time evolution factor
\begin{align}
  C_{m}(\tau^\ast,\tau)
  &= \exp\left[ -\frac{1}{2}\int_{\tau}^{\tau^\ast} \frac{d\tau'}{\tau'}\, \widetilde \gamma^{(A)}_{O}(\tau') \right] \notag \\
  & = \frac{\widetilde{\cal Z}_O^{(A)}(\tau^\ast)} {\widetilde{\cal Z}_O^{(A)}(\tau)} \equiv 
  \frac{R_A(\tau^\ast)}{R_O(\tau^\ast)} \frac{R_O(\tau)}{R_A(\tau)},
  \label{eq:Ctau}
\end{align}
can be obtained directly from correlator ratios.

It is straightforward to obtain $\widetilde \gamma_O$ as a function of the renormalized coupling. At a given flow time, the GF running coupling is defined through the energy density,
\begin{align}
  g_{\rm GF}^2(\tau) = \frac{128\pi^2}{3(N_c^2-1)}\,\tau^2\langle E(\tau)\rangle\,,  
  \label{eq:gf_coupling}
\end{align}
for $SU(N_c)$ \cite{Luscher:2010iy}. 
By eliminating $\tau$ between Eqs.~\eqref{eq:gamma_tau} and \eqref{eq:gf_coupling}, we can express the anomalous dimension as a function of the running coupling, which is more naturally defined in the massless limit.
The mass-independent RG anomalous dimension is obtained by taking the infinite-volume and $m_q\to0$ chiral limits. Subsequently, its continuum limit can be determined by taking the $a^2/\tau\to0$ limit at fixed $g^2_{\rm GF}$, analogous to the determination of the continuous GF $\beta$ function \cite{Fodor:2017die,Hasenfratz:2019hpg,Hasenfratz:2019hpg,Hasenfratz:2023nnp}.
The resulting function
\begin{align}
  \widetilde \gamma^{(A)}_{O}(g_{\rm GF}^2)\, .
\end{align}
 depends on the renormalization scheme (here the GF $A$ scheme), but is otherwise universal and independent of
the lattice action or the details of the flow. The anomalous dimension in the $V$ scheme $\widetilde \gamma^{(V)}_{O}(g_{\rm GF}^2)$ can be defined similarly.
Using the $V$ scheme definition, the anomalous dimensions of pseudoscalar and tensor operators have been investigated in Refs.~\cite{Carosso:2018bmz,Hasenfratz:2022wll,Hasenfratz:2023sqa,Hasenfratz:2023wbr}.

The evolution factor in the chiral limit,
\begin{equation}
    C_{\rm RG}(\tau^*,\tau) \equiv \lim_{m_q\to0}
    C_m(\tau^*,\tau)\,,    
\end{equation}
can be rewritten as
\begin{align}
  C_{\rm RG}(\tau^\ast,\tau)
  = \exp\left[ \frac{1}{2}\int_{g^2_\text{GF}(\tau)}^{g^2_\text{GF}(\tau^\ast)} dg^2\, \frac{\widetilde{\gamma}_{O}^{(A)}(g^2)}{\beta_{\rm GF}(g^2)} \right] ,
  \label{eq:Cg}
\end{align}
where the GF $\beta$ function is given by
\begin{align}
  \beta_{\rm GF}(g^2)
  = -\,\tau\frac{d g^2_{\rm GF}}{d\tau} \equiv \mu \, g_{\rm GF}\,\frac{d g_{\rm GF}}{d\mu}.
\end{align}
The evolution factor $C(\tau^\ast,\tau)$ as a function of $\tau$ or $g^2_\mathrm{GF}$ can be determined 
from lattice ensembles spanning the weak-to-strong coupling regime. 
The ratio form of $C_{m}(\tau^*,\tau)$ in Eq.~\eqref{eq:Ctau} can be determined directly from lattice correlators on physically large volumes at the same valence quark mass as the operator being renormalized, and its $a\to0$ continuum limit can be taken at fixed $\tau$ and $m_q$ (in physical units).
In contrast, Eq.~\eqref{eq:Cg} uses the massless RG functions $\widetilde\gamma_O^{(A)}(g_{\rm GF}^2)$ and $\beta_{\rm GF}(g_{\rm GF}^2)$. They can be calculated from lattice simulations performed in deconfined, small physical volumes in the chiral limit. The continuum predictions are obtained after taking the infinite-volume limit, followed by the $a\to 0$ continuum limit \cite{Fodor:2017die,Hasenfratz:2019puu,Hasenfratz:2023sqa,Hasenfratz:2023wbr}.

With the use of Eq.~\eqref{eq:Otau} to evolve lattice flow times $\tau$ to a more perturbative regime in $\tau^*$, the matching to \MSbar{} can be performed via
\begin{align}
  \langle O_{\overline{\rm MS}}(\mu^\ast) O(0) \rangle
  = \lim_{\tau^\ast\to0} & ~\widetilde{\zeta}_O^{-1}(\mu^\ast,\tau^\ast)\, \notag \\
  \times&~ C_m(\tau^\ast,\tau)\,
  \langle \widetilde{O}_{\rm GF}^{(A)}(\tau) O(0) \rangle\, .
    \label{eq:master_matching} 
\end{align}

The above relation connects the flowed quantity $\widetilde{O}^{(A)}_{\rm GF}(\tau)$ to the corresponding \MSbar{} observable. 
From lattice simulations, the operator $\widetilde{O}^{(A)}_{\rm GF}(\tau)$ is defined at finite lattice spacing $a$.
The flow-time evolution operator $C_m(\tau^*,\tau)$ in Eq.~\eqref{eq:master_matching} should be evaluated in the same GF scheme and at the same valence quark mass as the flowed operator. 
The final conversion to \MSbar{} uses the massless SFTX coefficient at the short flow time $\tau^\ast$, with finite-mass effects entering as higher-dimensional corrections suppressed by $\tau^\ast m_q^2$.
To allow $\tau^*$ to be as small as possible without increasing other discretization effects, $C_m(\tau^*,\tau)$ should be known in the continuum limit and thus can be universally determined for each quark mass and applied to any set of lattice data. 
Equation~\eqref{eq:master_matching} subsequently predicts results in the \MSbar{} scheme for arbitrarily high scales $\mu^*\propto1/\sqrt{\tau^*}$ without relying on perturbative  running.
Following the method of Ref.~\cite{Black:2025gft}, one would take the continuum limit of $\langle \widetilde{O}_{\rm GF}^{(A)}(\tau) O(0) \rangle$ before applying Eq.~\eqref{eq:master_matching}.
Alternatively, if the SFTX matching and the anomalous dimension do not mix with other operators (i.e.~are scalar quantities), then Eq.~\eqref{eq:master_matching} is a linear relation, and the continuum limit can be taken after the $\tau^*\to0$ limit.
Determining $\langle O_{\overline{\rm MS}}(\mu^\ast) O(0) \rangle$ for multiple lattice spacings before taking its continuum limit is more similar to other renormalization methods. We will explore the advantages of this approach in Section~\ref{sec:quarkmasses}.

\section{Applications}
\label{Sec.Applications}
We now illustrate the GF $A$ and $V$ scheme constructions for a small set of applications using flowed two-point correlation functions generated on RBC-UKQCD's 2+1 flavor Shamir domain-wall fermion ensembles~\cite{Allton:2008pn,RBC:2010qam,RBC:2014ntl,Boyle:2017jwu,Boyle:2018knm}. Details are summarized in Table \ref{Tab.Ensembles}.
We first use the zero-flow-time limit of $R_A/R_V$ to test the relation between the Ward-identity normalizations $Z_A$ and $Z_V$. 
We next determine the finite-mass flowed anomalous dimension $\widetilde\gamma_m^{(A)}(\tau)$ along with the corresponding evolution factor $C_m(\tau^\ast,\tau)$, and use these together with $Z_\chi^{(A)}(\tau)$ to determine the renormalized strange quark mass in the \MSbar{} scheme. 
Unless stated otherwise, the correlators are connected $s\bar{s}$ correlators with valence strange quark masses tuned at or near their physical values.
The correlation functions \cite{Black:2026data} used below were originally generated for the projects of Refs.~\cite{Black:2025gft,Black:2026dzp, Black:2026rbz}.   

\subsection{Ratio \texorpdfstring{$Z_V/Z_A$}{ZV/ZA}}

The ratio of GF-renormalized operators in different
schemes  determines the ratio of the corresponding
wavefunction renormalization factors, i.e.
\begin{align}
  \frac{\widetilde{O}_{\mathrm{GF}}^{(A)}(\tau)}{\widetilde{O}_{\mathrm{GF}}^{(V)}(\tau)}
  = \frac{\widetilde{\cal Z}_\chi^{(A)}(\tau)}{\widetilde{\cal Z}_\chi^{(V)}(\tau)} = \frac{Z_A}{Z_V} \frac{R_A(\tau)}{R_V(\tau)} \, 
  \label{eq:O_ratio}
\end{align}
where we have used Eq.~\eqref{eq:ZchiA} and its analogue for $V$.
This ratio is finite and dimensionless; consequently, the dependence on the flow time must manifest itself through the only dimensionless combination available, $\tau m_q^2$, where $m_q$ is the bare quark mass used to calculate the lattice two-point correlation functions. 
In the limit $\tau \to 0$, the mass dependence is suppressed and  $\widetilde{\cal Z}_\chi^{(A)}(0) = \widetilde{\cal Z}_\chi^{(V)}(0)$, while at finite $\tau$
\begin{align}
  \begin{aligned}
    \frac{\widetilde{\cal Z}_\chi^{(A)}(\tau)}{\widetilde{\cal Z}_\chi^{(V)}(\tau)} 
    &= 1 + \mathcal O(\tau m_q^2) \, .
  \end{aligned}
  \label{eq:tildeZAZV}
\end{align}
Thus in the $\tau\to 0$ limit we have
\begin{align}
  \frac{Z_V}{Z_A} = \lim_{\tau \to 0} \frac{R_A(\tau)}{R_V(\tau)} \, .
  \label{eq:ZVZA}
\end{align}
The ratio $Z_V/Z_A$ is independent of the GF; it characterizes the lattice operators and depends on the specific choice of lattice action employed, as well as the lattice spacing. 
Equation \eqref{eq:ZVZA} shows how this ratio can be predicted based on GF correlators in the $\tau \to 0$ limit.

\begin{figure}[tb]
  \includegraphics[width=0.48\textwidth]{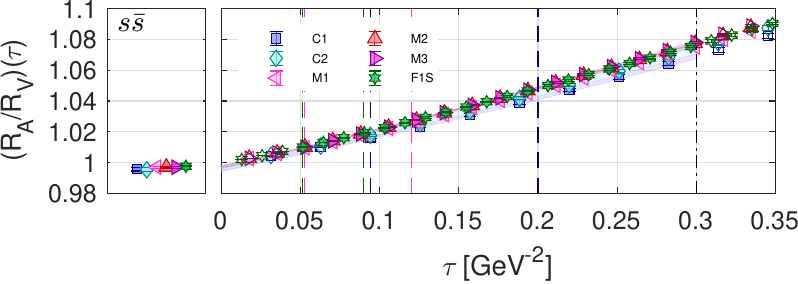}
  \caption{$R_A(\tau)/R_V(\tau) = \widetilde{\cal Z}_\chi^{(A)}(\tau) /\widetilde{\cal Z}_\chi^{(V)}(\tau) \times (Z_V/Z_A)$ vs.~$\tau$ using connected $\bar s s$ correlators.}
  \label{Fig.RatioZchiAZchiV_vs_tau_GeV2}
\end{figure}

We demonstrate the $\tau$ dependence of $(R_A/R_V)(\tau)$ in Fig.~\ref{Fig.RatioZchiAZchiV_vs_tau_GeV2}. 
To extract the value at zero flow time, we perform the $\tau \to 0$ extrapolation procedure described in Ref.~\cite{Black:2025gft} and summarized in Appendix \ref{Sec.ZeroFlowTime}. 
For this analysis, we vary $2/(8 a^{-2}) \le \tau_\text{min} \le 5/(8a^{-2})$ with fixed $\tau_\text{max}= 0.3\,\text{GeV}^{-2}$ as well as $0.25\,\text{GeV}^{-2} \le\tau_\text{max} \le 0.35\,\text{GeV}^{-2}$ for fixed $\tau_\text{min}=4/(8a^{-2})$. 
We report the result at the strange quark mass using $(s\bar s)$-correlators in the second column of Table \ref{Tab.ZchiAZchiV}.
The third column of Table \ref{Tab.ZchiAZchiV} shows the ratios $Z_V/Z_A$ obtained by using RBC-UKQCD determinations at the unitary light quark mass (for details and references see Table \ref{tab:Z_A-V}). 
As a consistency check, we show in the fourth column the product of $[R_A/R_V]_{\tau=0}$ (obtained with $s\bar s$ correlators) with the available unitary-light-quark determinations of $Z_V/Z_A$. 
When $Z_V/Z_A$ is evaluated for the same valence quark mass as $[R_A/R_V]_{\tau=0}$, the product should be unity; using the available results instead introduces a small mass-dependent systematic.

\begin{table}[t]
  \begin{tabular}{cccc}
    \hline\hline
    & \begin{minipage}{0.28\columnwidth}$$ \left[\frac{R_A}{R_V}\right]^{s\bar s}_{\tau=0}$$\end{minipage}
    & \begin{minipage}{0.28\columnwidth}$$
    \left[\frac{Z_V}{Z_A}\right]^{\ell\bar \ell}
    $$\end{minipage}
    & \begin{minipage}{0.28\columnwidth}$$ \left[\frac{\widetilde{\cal Z}_\chi^A}{\widetilde{\cal Z}_\chi^V}\right]^{s\bar s,\ell\bar\ell}_{\tau=0}$$\end{minipage}\\[6mm] \hline
    C1 &0.9957(11)\phantom{0}& 0.99548(84) &1.0004(18)\\
    C2 &0.9953(12)\phantom{0}& 0.99479(35) &1.0003(14)\\
    M1 &0.99704(61)& 0.9995(13)\phantom{0} &0.9979(21)\\
    M2 &0.99707(59)& 0.99819(74) &0.9986(13)\\
    M3 &0.99694(62)& 0.99869(57) &0.9979(12)\\
    F1S &0.99758(24)& --- &\textit{1.0004(10)}\\
    \hline\hline
  \end{tabular}
  \caption{Ratios $R_A/R_V$ predicting $Z_V/Z_A$ from $(s\bar s)$ correlators obtained for all six ensembles. The third column shows the values of $Z_V/Z_A$ as determined by RBC-UKQCD at the unitary light quark mass (for details see Table \ref{tab:Z_A-V}). The fourth column shows the consistency check $[(Z_A R_A)/(Z_VR_V)]_{\tau=0} \equiv [\widetilde{\cal Z}_\chi^A/\widetilde{\cal Z}_\chi^V]_{\tau=0} \simeq 1$. The value for F1S is set in italics because $Z_V$ has not been independently determined.}
\label{Tab.ZchiAZchiV}
\end{table}

\subsection{Numerical determination of \texorpdfstring{$\widetilde \gamma_O^{(A)}$ and $C_m(\tau^*,\tau)$}{gammaO and Cm(tau*,tau)}}

\begin{figure}[tp]
  \includegraphics[width=0.48\textwidth]{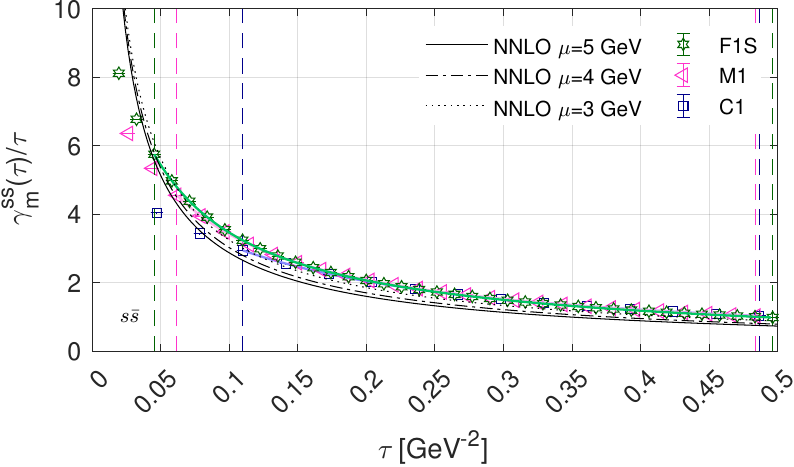}\\
  \caption{$\widetilde\gamma^{(A)}_m/\tau$ vs.~$\tau$ for three of the ensembles with $a^{-1}= 2.78$, 2.38, and 1.78 GeV. The dashed vertical lines with matching colors indicate $\tau/a^2=2/8$.}
  \label{Fig.gamma}
\end{figure}

Following the expression given in Eq.~\eqref{eq:gamma_tau}, $\widetilde \gamma_O^{(A)}(\tau)$ can be determined from the ratios of two-point functions. Figure~\ref{Fig.gamma} shows $\widetilde \gamma_O^{(A)}(\tau)/\tau$ as a function of $\tau$ for three of our ensembles with lattice spacings $a^{-1}= 2.78$, 2.38, and 1.78 GeV, obtained using correlators with mass at or near the physical strange quark mass.
Despite the small difference in the strange quark mass on F1S w.r.t.~the medium and coarse ensembles, as well as different values of the lattice spacing, the predictions are consistent at sufficiently large flow times.
Once the flow time in lattice units reaches $\tau/a^2 \gtrsim 2/8$ (indicated by the left dashed vertical lines matching the color of the corresponding data points), the $a \to 0$ continuum limit can be taken. 
For values of $\tau/a^2 \lesssim 2/8$, the data show significant cutoff effects, which is consistent with the GF smearing radius being too small relative to the lattice spacing.
Based on these observations, we use data at our finest lattice spacing (ensemble F1S) to predict  $\widetilde \gamma_O^{(A)}(\tau)/\tau$ at the strange quark mass for $\tau \gtrsim 0.045\, \text{GeV}^{-2}$ with only small $\mathcal{O}(a^2)$ cutoff effects. 
In practice, we perform cubic spline interpolations for $2/(8a^{-2})< \tau < 0.5\,\text{GeV}^{-2}$ and obtain an error band for $\gamma_m(\tau)/\tau$ by interpolating our data, coherently shifting the values up/down by $\pm 1\sigma$. 
Looking towards the future, simulations with half the lattice spacing ($a^{-1} \approx 5.5\, \text{GeV}^{-1}$) would already extend the accessible flow time range down to $\tau \gtrsim 0.01\,\text{GeV}^{-2}$.

\begin{figure}[tp]
  \includegraphics[width=0.48\textwidth]{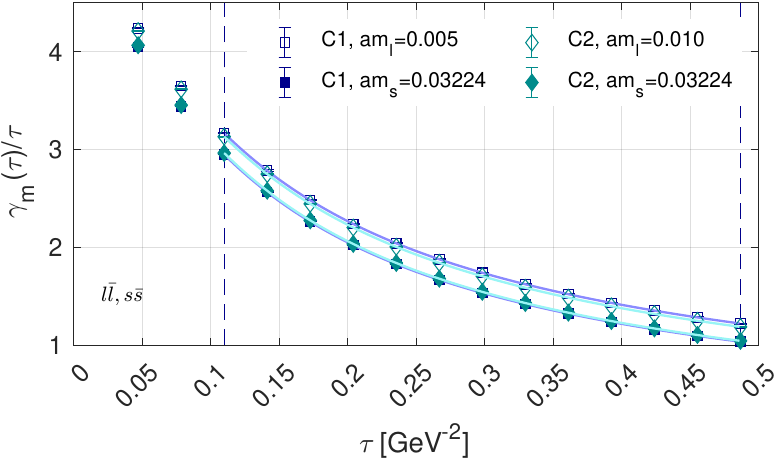}
  \caption{ Dependence of $\widetilde\gamma^{(A)}_m/\tau$ vs.~$\tau$ on the valence quark mass on the coarse C1 and C2 ensembles. Compared to Fig.~\ref{Fig.gamma}, the vertical axis shows a much smaller range.}
  \label{Fig.gamma_ll_ss}
\end{figure}

In the typical RG picture, the anomalous dimension of an operator is defined in the massless limit. However, in Fig.~\ref{Fig.gamma} we show $\widetilde \gamma_O^{(A)}(\tau)$ as defined in Eq.~\eqref{eq:gamma_tau} along a line of constant physics with finite mass. This definition clearly depends on the fermion mass; this mass dependence is relatively small and is suppressed as the flow time $\tau \to 0$. 
As a demonstration, in Fig.~\ref{Fig.gamma_ll_ss} we show $\widetilde{\gamma}_m(\tau)/\tau$ obtained on the coarse C1 and C2 ensembles, using the unitary light quark mass as well as the physical strange quark mass.
While the difference between the unitary light quark masses on the two ensembles is barely resolved, the physical strange quark mass leads to a small but consistent downward shift.

At small $\tau$, the nonperturbative expression should match the prediction from the SFTX. The SFTX is a perturbative expansion in terms of the strong coupling $\alpha_s(\mu)$ and depends on the combination $\mu^2 \tau$ logarithmically. Thus, at any finite order, the predicted $\widetilde\gamma_O(\tau)$ has a remnant $\mu$ dependence. 
In Fig.~\ref{Fig.gamma}, we compare the NNLO predictions at $\mu=3$, 4, and 5 GeV (dotted, dash-dotted, solid lines) to the nonperturbative values. At small flow time, $\tau \lesssim 0.02\,\text{GeV}^{-2}$, the $\mu$ dependence diminishes. In the $0.045 \lesssim \tau \lesssim 0.075\, \text{GeV}^{-2}$ region, we observe close agreement between the SFTX and nonperturbative predictions. 
We therefore approximate $C_m(\tau^*,\tau)$ by using the perturbative NNLO result for  $0\le \tau \le 0.05\,{\rm GeV}^{-2}$, where we connect it to our nonperturbative determination obtained from our data for the strange quark mass on the F1S ensemble. 

\subsection{Renormalized strange quark mass}\label{sec:quarkmasses}
In this section we demonstrate how to determine the renormalized strange quark mass from GF quantities. In analogy to continuum calculations, we define a renormalized quark mass in the GF scheme as
\begin{equation}
\widetilde m^{(A)}_{GF}(\tau;a) =  \frac{m_{bare}(a)}{\widetilde Z_P^{(A)}(\tau;a)}. 
\end{equation}
such that $\widetilde m_{GF}^{(A)}$ is obtained for each ensemble at finite lattice spacing $a$ and $\widetilde Z_P^{(A)} = Z_A R_A/R_P$. 
For domain-wall fermions the bare input quark mass is given by
\begin{align}
  a m_\text{bare} = am_0 + am_\text{res},
\end{align}
with $am_0$ the input quark mass and $am_\text{res}$ parameterizing the residual chiral-symmetry breaking due to the finite extent of the fifth dimension $L_s$. 
Matching  $\widetilde m_{GF}^{(A)}$ to the \MSbar{} scheme via the SFTX is given by
\begin{align}
  m_{\msbar}(\mu_{UV},a)
  = \lim_{\tau \to 0} \, \zeta^{-1}_{AP}(\mu_{UV},\tau^*)\, C_m(\tau^\ast,\tau) \widetilde m_{GF}^{(A)}(\tau;a).
  \label{Eq.mMSbar}
\end{align}
The SFTX matching coefficient $\zeta_{AP}^{-1} = \zeta_A^{-1}\zeta_P$ depends on the renormalization scale $\mu_{UV}$ and the flow time $\tau$. The number of active flavors ($N_f=4$) and the values of $\alpha_s(\mu)$ also enter its determination. 
When the relevant flowed observable is combined with its SFTX matching coefficient, the \MSbar{} result is obtained by taking the $\tau\to0$ limit.
In order to suppress large logarithms in the matching coefficients, the flow time $\tau^*$ can be chosen such that $\mu_{UV}^2 \tau^\ast \approx 1$.
The lattice data are evolved from $\tau$ to $\tau^*$ via $C_m(\tau^\ast,\tau)$, which uses $\widetilde\gamma_m^{ss}(\tau)$. 
The anomalous dimension is determined at the strange quark mass on the F1S ensemble for the nonperturbative regime of the evolution, and we connect it at $\tau_\text{con} = 0.05\,\text{GeV}^{-2}$ to the NNLO massless perturbative prediction, where finite-mass corrections are suppressed by powers of $\tau m_s^2$ ($\lesssim0.05\%$).
For the perturbative part of $\widetilde\gamma_m(\tau)$, we choose the same $\mu_{UV}$ as for $\zeta_{AP}^{-1}$. 

\begin{figure}[t]
  \includegraphics[width=0.48\textwidth]{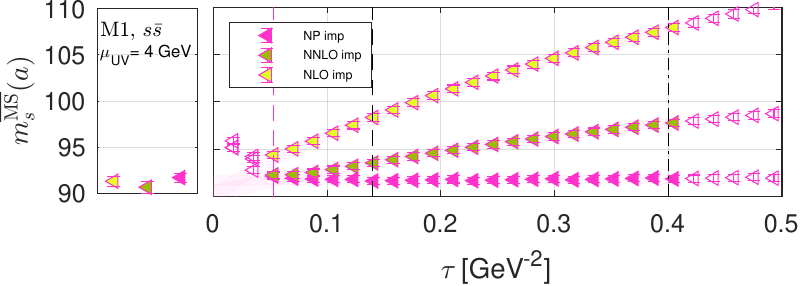}
  \caption{Using the renormalized strange quark mass for the M1 ensemble as an example, we compare the RG improvement obtained with our nonperturbatively-improved $\zeta_{AP}^{-1}$ (pink filled symbols) to perturbatively-improved values at NLO (yellow filled symbols) and NNLO (green filled symbols). 
  } 
  \label{Fig.CompareTauExtra}
\end{figure}

In Fig.~\ref{Fig.CompareTauExtra} we demonstrate the nonperturbative improvement of $\zeta_{AP}^{-1}$ by comparing determinations of the renormalized strange quark mass on the M1 ensemble. 
The pink filled triangles show our new results using the nonperturbative anomalous dimension $\widetilde \gamma_m(\tau)/\tau$. The $\tau$ dependence is almost entirely removed, and the $\tau\to0$ extrapolation is very flat and is, therefore, based on a linear ansatz. 
Also shown are results using perturbatively-improved $\zeta_{AP}^{-1}$~\cite{Artz:2019bpr,Borgulat:2025gys} at NLO (yellow filled symbols) and NNLO (green filled symbols) which are used in the analysis of Ref.~\cite{Black:2025gft}. To account for remaining logarithmic corrections, we perform the $\tau\to 0$ extrapolation using the ansatz $f_l(\tau) = \tau (c_l \log(\tau\mu^2) + c_1) + c_0$. 
While perturbatively-calculated/improved matching coefficients $\zeta^{-1}_{AP}$ are valid only in the small flow-time region, the nonperturbative evolution can be employed at arbitrarily large flow times without an upper limit to match a larger range of lattice data to perturbatively-compatible flow times.

\begin{figure}[tb]
\includegraphics[width=\columnwidth]{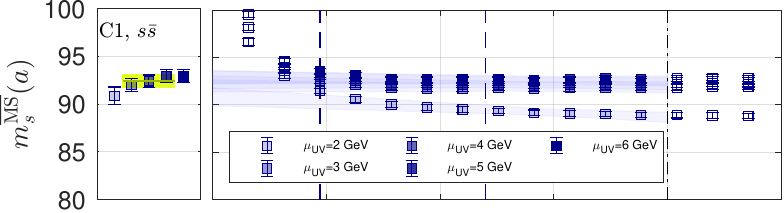}\\ 
\includegraphics[width=\columnwidth]{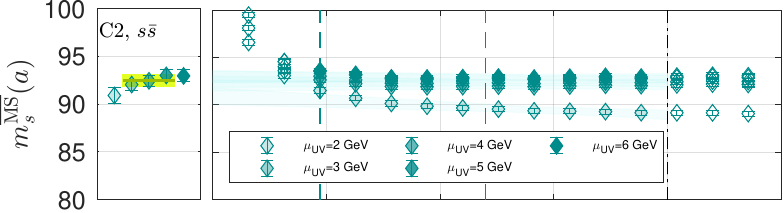}\\
\includegraphics[width=\columnwidth]{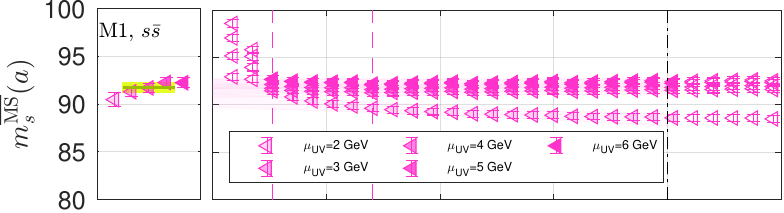}\\ 
\includegraphics[width=\columnwidth]{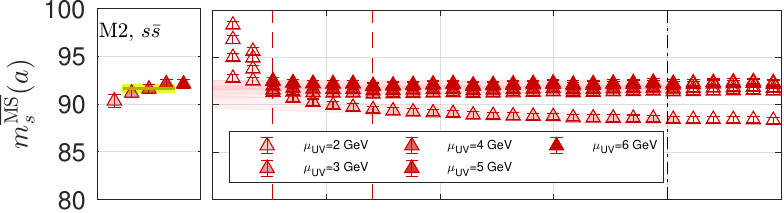}\\
\includegraphics[width=\columnwidth]{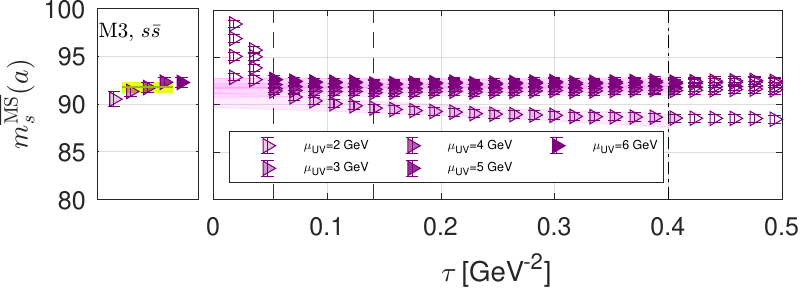}
\caption{Determination of the strange quark mass for the C1, C2 and M1, M2, M3 ensembles using the $A$-flow scheme and nonperturbative $\gamma_m$ determined on the F1S ensemble. In Eq.~\eqref{Eq.mMSbar} we use $\mu_\text{UV}=2,\,3,\,4,\,5,\,6\,\text{GeV}$ and 4-loop \MSbar{} running \cite{Chetyrkin:2000yt,Herren:2017osy} to always show values at the reference scale $\mu=2\,\text{GeV}$. The $\tau\to 0$ extrapolation is explained in App.~\ref{Sec.ZeroFlowTime}.} 
\label{Fig.ms}
\end{figure}

\begin{figure}[tb]
\includegraphics[width=\columnwidth]{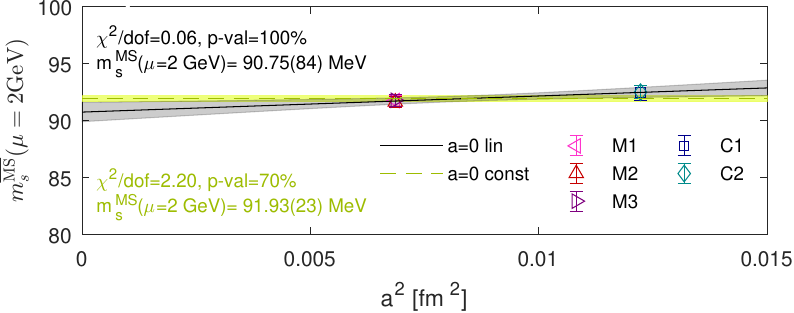}
\caption{Continuum limit extrapolation to obtain the strange quark mass in the \MSbar{} at the reference scale $\mu=2\,\text{GeV}$ using the correlated averages for $\mu_{UV}=3,\,4,\,5\,\text{GeV}$ presented in Fig.~\ref{Fig.ms}.}
\label{Fig.msCont}
\end{figure}

In Fig.~\ref{Fig.ms} we show the determination of the \MSbar{} strange quark mass for all five medium and coarse ensembles. 
We vary $\mu_{UV}$ from 2 to 6 GeV and use 4-loop \MSbar{} running \cite{Chetyrkin:2000yt,Herren:2017osy} to obtain the \MSbar{} values at the reference scale of $\mu=2\,\text{GeV}$. 
As can be seen in Fig.~\ref{Fig.ms}, using $\mu_{UV}=2\,\text{GeV}$ causes a significant downward shift of the data, indicating that $\mu_{UV}=2\,\text{GeV}$ is likely too small. 
We perform the $\tau \to 0$ limit as described in App.~\ref{Sec.ZeroFlowTime}, using only flow times $\tau/a^2\ge 2/8$. 
Subsequently, for each ensemble, we calculate a single correlated average~\cite{Schmelling:1994pz} of the results for $\mu=3,\,4,$ and $5\,\text{GeV}$.
These values for each ensemble feed into the final step of taking the $a\to 0$ continuum limit, as shown in Fig.~\ref{Fig.msCont}. 
Since our ensembles feature only two different lattice spacings, we perform both linear-in-$a^2$ and constant-in-$a^2$ continuum limit extrapolations.
Both extrapolations have excellent $p$-values, and the results agree within uncertainties when accounting for additional systematic effects. Calculating the correlated average of both values, we obtain for the renormalized strange quark mass in the \MSbar{} scheme at $\mu=2\,\text{GeV}$
  \begin{align} \label{eq:msmsbar}
    m_s^{\msbar}(\mu=2\,\text{GeV}) = 91.8(4)\,\text{GeV},
  \end{align}
where the error accounts for the statistical uncertainty as well as part of the $\tau\to 0$ extrapolation.

\begin{figure}[t]
  \includegraphics[width=0.48\textwidth]{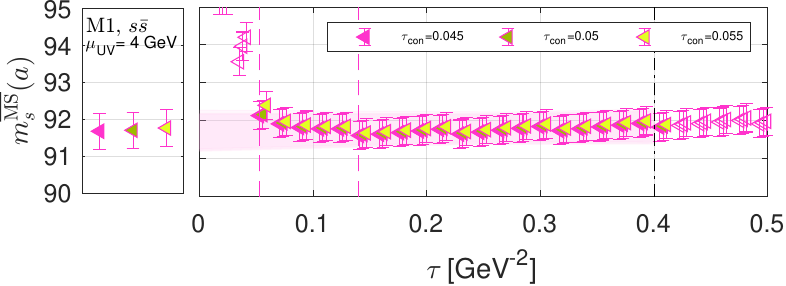}
  \caption{Comparison of the renormalized strange quark mass on the M1 ensemble determined using different  flow times $\tau_{\rm con}$  to connect the nonperturbative  to the perturbative anomalous dimensions. The central value $\tau_{\rm con}=0.05\,{\rm GeV}^{-2}$ is used for the pink-filled data, while the yellow-filled and green-filled data show $\tau_{\rm con}\pm10\%$ respectively.}
  \label{Fig.VaryTauCon}
\end{figure}

Other systematic effects arise, e.g., from the flow time $\tau_\text{con}$ at which we connect the nonperturbatively-improved anomalous dimension to that of the perturbative calculation. By varying $\tau_\text{con}$, we can check for the significance of this effect. As shown in Fig. \ref{Fig.VaryTauCon}, changing $\tau_\text{con}=0.05$ to 0.045 or 0.055 has almost no impact on our determination and, hence, is a negligible effect. 
We note, however, that the nonperturbative data used for $\gamma_m(\tau)$ are approximated by the measurements on the F1S ensemble only, without performing a continuum extrapolation. As this introduces a systematic uncertainty that is hard to quantify with our dataset, we do not provide a full error budget.  
The analysis shown here is only demonstrative. 

\smallskip
The analysis to obtain renormalized quark masses using a GF scheme is slightly special because the SFTX matching coefficient $\zeta_{AP}^{-1}$ is scalar. Hence the order of the $\tau\to 0$ and $a\to 0$ limits can be interchanged. In contrast to our prior work in Ref.~\cite{Black:2025gft}, here we choose to first take the $\tau\to0$ limit which is very similar to the procedure conventionally used with renormalization coefficients calculated perturbatively or using e.g.~an RI/MOM scheme.

\section{Conclusion}
\label{Sec.Conclusion}
We have introduced GF renormalization prescriptions for fermionic composite operators, in which the fermion wavefunction renormalization $\widetilde{\cal Z}_\chi(\tau)$ is fixed nonperturbatively by either the partially conserved axial charge or the conserved vector current. The resulting $A$ and $V$ schemes are defined entirely through flowed two-point correlation functions. They avoid backward-flow constructions required by local ringed-scheme definitions, while retaining a direct connection to the short-flow-time expansion. In the short-flow-time limit, both $A$ and $V$ schemes can be matched to \MSbar{} using the known ringed-scheme SFTX matching coefficients.

Analogous correlator ratios define GF renormalization factors for general composite fermionic operators. The logarithmic derivatives of these ratios with respect to the flow time yield nonperturbative anomalous dimensions which, in turn, define the evolution factor 
$C(\tau^\ast,\tau)$ that transports operators from lattice-accessible flow times to shorter flow times, where perturbative SFTX matching is more reliable.
This framework thus decomposes GF renormalization into a nonperturbative flow-time evolution stage followed by a final perturbative matching step to the \MSbar{} scheme.

We illustrated the proposed construction using RBC-UKQCD's Shamir domain-wall fermion ensembles and existing flowed meson correlators. As a first application we determined the ratio of the matching factors $Z_V/Z_A$ and showed that the results agree with independent lattice determinations. 
We also determined the finite-mass flow-time anomalous dimension of the pseudoscalar density -- equivalent to the quark mass anomalous dimension -- near the physical strange quark mass. The corresponding evolution factor removes most of the residual flow-time dependence of the GF-renormalized strange quark mass, stabilizing the $\tau\to0$ extrapolation and improving the connection to \MSbar{}. 
Our determination is in good agreement with our previous results in Ref.~\cite{Black:2025gft}, 
and with the FLAG averages~\cite{FlavourLatticeAveragingGroupFLAG:2024oxs} for 2+1~\cite{CLQCD:2023sdb,Bruno:2019vup, RBC:2014ntl, McNeile:2010ji, BMW:2010ucx,BMW:2010skj, MILC:2009ltw} 
and 2+1+1~\cite{ExtendedTwistedMass:2021gbo,Lytle:2018evc,FermilabLattice:2018est,Chakraborty:2014aca,EuropeanTwistedMass:2014osg} flavors 
as well as a 
new 2+1 flavor result \cite{DelDebbio:2024hca} not  yet included in the average. 
While we expect our systematic uncertainties to be competitive, quantifying the impact of approximating $\widetilde\gamma_m(\tau)$ by using F1S data instead of a continuum extrapolated result is troublesome and we refrain from quoting a full error budget.

The numerical analysis presented here is intended as a demonstration of the method. A universal anomalous dimension $\widetilde\gamma_O(g^2_{\rm GF})$ requires the chiral and infinite-volume limits, followed by a continuum extrapolation at fixed GF coupling to predict the flow time evolution factor $C_{\rm RG}(\tau^\ast,\tau)$. 
The finite-mass evolution factor $C_m(\tau^\ast,\tau)$ is appropriate to match large-volume numerical data at finite mass. Its $a\to 0$ continuum limit is taken at fixed flow time.  As the flow time decreases, it matches the massless evolution factor and eventually the SFTX-predicted perturbative evolution. 
Future work should extend the calculation of $C_m(\tau^*,\tau)$ to additional lattice spacings, lighter and also heavier valence masses, and operators with non-scalar matching coefficients and anomalous dimensions; also a complete calculation of $C_{\rm RG}(\tau^*, \tau)$ is desirable.

Further applications include bag parameters, mixed-action current normalizations, and other matrix elements requiring absolute operator normalization. More broadly, the evolution factor $C_{m,{\rm RG}}(\tau^\ast,\tau)$ provides a systematic bridge between lattice flow times, where cutoff effects are controlled, and short flow times, where perturbative matching to continuum schemes is applicable.

\section*{Acknowledgments}
We thank the RBC/UKQCD Collaboration for generating and making their
gauge field ensembles publicly available. The data analyzed here were generated as part of the project to calculate heavy meson lifetimes~\cite{Black:2026rbz,Black:2026dzp}. 

We thank Robert Harlander, Jonas Kohnen, Fabian Lange, and Antonio Rago for fruitful discussions, including those that took place during the Higgs Centre workshop: ``Standard Model parameters and observables from gradient flow'' in Edinburgh, Scotland, May 12$^{\rm th}$--15$^{\rm th}$, 2026.
A.H.~and O.W.~are grateful for the hospitality at the Kavli Institute of Theoretical Physics at UC Santa Barbara, where part of this work was carried out during the program “What is Particle Theory?”.

These computations used resources provided by the OMNI cluster at the University of Siegen, the HAWK cluster at the High-Performance Computing Center Stuttgart, and LUMI-G at the CSC data center Finland (DeiC National HPC g.a.~DEIC-SDU-L5-13 and DEIC-SDU-N5-2024053).

A.H.~acknowledges support from DOE grant DE-SC001000.
M.B.~was funded by UK STFC grant ST/X000494/1.
O.W.~received support from the Deutsche Forschungsgemeinschaft (DFG, German Research Foundation) through grant 396021762 — TRR 257 ``Particle Physics Phenomenology after the Higgs Discovery'' and from the Deutsche Forschungsgemeinschaft (DFG, German Research Foundation) under Germany’s Excellence Strategy -- EXC 3107 -- Project 533766364.
This research was supported in part by grant NSF PHY-2309135 to the Kavli Institute for Theoretical Physics (KITP).

\section*{Data availability}
The data that support the findings of this article are openly available~\cite{Black:2026data}.

\appendix
\section{Numerical setup}
\label{Sec.Setup}
We illustrate our method using \mbox{RBC-UKQCD}'s 2+1 flavor Shamir domain wall fermion ensembles \cite{Allton:2008pn,RBC:2010qam,RBC:2014ntl,Boyle:2017jwu,Boyle:2018knm}.  In particular, we analyze the coarse C1 and C2 ensembles, the M1, M2, and M3 medium ensembles, and the F1S fine ensemble.
The corresponding lattice volumes and spacings, residual masses, and input bare quark masses are summarized in Table~\ref{Tab.Ensembles}.

\begin{table*}[t]
  \centering
  \begin{tabular}{l@{~~~}c@{~~~}c@{~~~}c@{~~~}c@{~~~}c@{~~~}cccccc}
    \hline\hline & L/a & T/a & $a^{-1}[\text{GeV}]$ & $M_\pi[\text{MeV}]$ & $am_\text{res}$ &$am_\ell^\text{sea}$ & $am_s^\text{sea}$ & $am_s^\text{val}$ & $N_\text{src}\times\text{N}_{\text{conf}}$ \\\hline\hline
    C1 & 24 & 64 & 1.7848(50) & 339.8(1.2) & 0.003154(15)\phantom{0} &0.005\phantom{144} & 0.04\phantom{144} & 0.03224 & $32\times101$ \\
    C2 & 24 & 64 & 1.7848(50) & 430.6(1.4) & 0.003154(15)\phantom{0} &0.010\phantom{144} & 0.04\phantom{144} & 0.03224 & $32\times101$ \\
    M1 & 32 & 64 & 2.3833(86) & 303.6(1.4) & 0.0006697(34)&0.004\phantom{144} & 0.03\phantom{144} & 0.02477 & $32\times\phantom{0}79$  \\
    M2 & 32 & 64 & 2.3833(86) & 360.7(1.6) & 0.0006697(34)&0.006\phantom{144} & 0.03\phantom{144} & 0.02477 & $32\times\phantom{0}89$\\
    M3 & 32 & 64 & 2.3833(86) & 410.8(1.7) & 0.0006697(34)&0.008\phantom{144} & 0.03\phantom{144} & 0.02477 & $32\times\phantom{0}68$\\
    F1S & 48 & 96 & 2.785(11) & 267.6(1.3) & 0.0009679(21)&0.002144 & 0.02144 & 0.02167 & $24\times\phantom{0}98$ 
    \\\hline\hline 
  \end{tabular}
  \caption{
    RBC/UKQCD's $N_f=2+1$ Shamir DWF ensembles with Iwasaki gauge action~\cite{Allton:2008pn,RBC:2010qam,RBC:2014ntl,Boyle:2017jwu,Boyle:2018knm} specified by the inverse lattice spacing ($a^{-1}$), unitary pion mass ($M_\pi$), $am_\text{res}$ parametrizing residual chiral symmetry breaking, light and strange sea quark masses ($am_\ell^\text{sea}$ and $am_s^\text{sea}$), and (near-)physical-mass valence strange ($am_s^\text{val}$) quark masses. 
    For the coarse (C1, C2), medium (M1, M2, M3), and fine ensemble (F1S) we use
    $N_\text{conf}$ number of configurations and place 
    $N_\text{src}$ evenly-spaced $\mathbb{Z}_2$
    wall-sources per configuration.
    }
  \label{Tab.Ensembles}
\end{table*}

The fermion flowed meson correlators analyzed in this work were generated as part of the heavy-meson lifetime calculation~\cite{Black:2026rbz,Black:2026dzp} and are also used to determine the strange and charm quark mass in Ref.~\cite{Black:2025gft}.

\begin{table*}[t]
\begin{tabular}{l@{~~~}l@{~~~}lc@{~~~~}lcl@{~~~~}}
\hline\hline
& $am_q$& $Z_A(am_q)$ & &$Z_V(am_q)$ &  & $Z_V/Z_A$\\
\hline
 C1 & 0.005 & 0.71732(14)&\cite{RBC:2010qam}  & 0.71408(58) &\cite{RBC:2014ntl}&0.99548(84)\\
    & 0.005 & 0.717247(67)&\cite{Boyle:2017jwu} & &\\
    & 0.005 & 0.71796(44) &\cite{DelDebbio:2024hca}& &\\
    & 0.03224 & 0.721079(48) & &\multicolumn{3}{c}{our determination}\\
    \hline
 C2 & 0.010 & 0.71783(15)&\cite{RBC:2010qam} &  0.71409(20)&\cite{RBC:2014ntl}&0.99479(35)\\
    & 0.010 & 0.717831(53)&\cite{Boyle:2017jwu} & &\\
    & 0.03224 & 0.721025(48) & &\multicolumn{3}{c}{our determination}\\
    \hline
 M1 & 0.004 & 0.745053(54)&\cite{RBC:2010qam} & 0.74470(99) &\cite{RBC:2014ntl}&0.9995(13)\\
    & 0.004 &0.744949(39) &\cite{Boyle:2017jwu}&&\\
    & 0.004 & 0.74486(27)  &\cite{DelDebbio:2024hca}& &\\
    & 0.02477 & 0.747372(36) &&\multicolumn{3}{c}{our determination}\\
    \hline
 M2 & 0.006 & 0.745222(45) &\cite{RBC:2010qam}& 0.74387(55)&\cite{RBC:2014ntl}&0.99819(74)\\
    & 0.006 & 0.745190(40) &\cite{Boyle:2017jwu}& &\\
    \hline
 M3 & 0.008 & 0.745328(48) &\cite{RBC:2010qam}&0.74435(42) &\cite{RBC:2014ntl}&0.99869(57)\\
    \hline
 F1S  & 0.0021 & 0.76231(18) &\cite{DelDebbio:2024hca}& \textit{0.76007(61)}&& ---\\
    & 0.0214 & 0.764180(44) &\cite{DelDebbio:2024hca}& &&\\
    & 0.02167 & 0.764260(15) &&\multicolumn{3}{c}{our determination}\\
 \hline\hline
\end{tabular}
\caption{Determinations of $Z_A$ using 5d conserved currents 
as well as $Z_V$ determinations based on pseudoscalar-vector-pseudoscalar 3-point functions. The value of $Z_V$ set in italics for the F1S ensemble is deduced from $Z_A$ and the ratio $Z_V/Z_A$ obtained using Eq.~\eqref{eq:ZVZA} (cf.~Table \ref{Tab.ZchiAZchiV}). } 
\label{tab:Z_A-V}
\end{table*}

\begin{figure}[t]
  \includegraphics[width=\columnwidth]{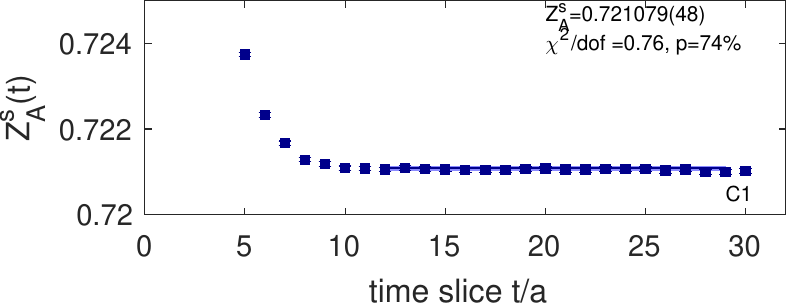}\\
  \includegraphics[width=\columnwidth]{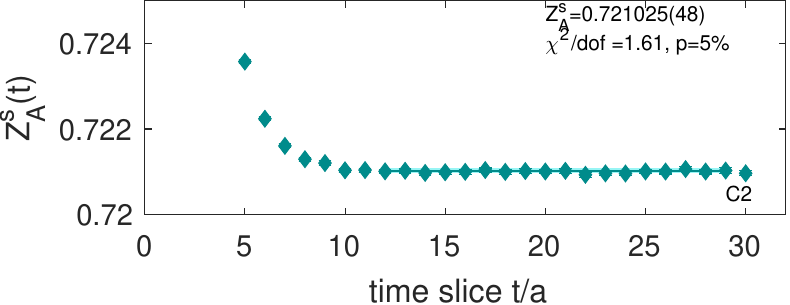}\\  
  \includegraphics[width=\columnwidth]{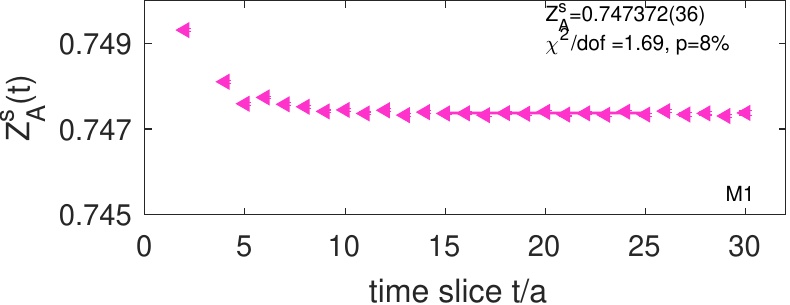}\\
  \includegraphics[width=\columnwidth]{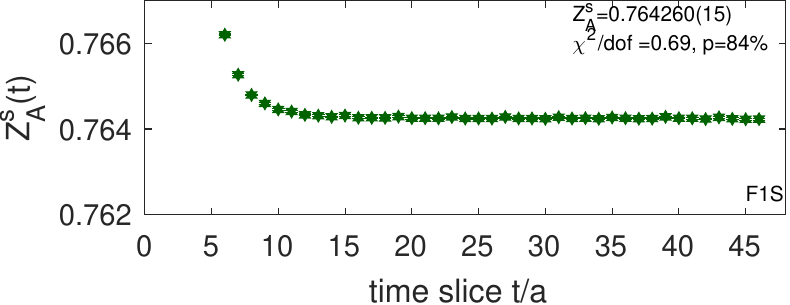}
  \caption{Determination of $Z_A$ using 5-d conserved currents for strange quark mass correlators.}
  \label{Fig.ZAs}
\end{figure}

In Table \ref{tab:Z_A-V} we collect the published values for the matching factors $Z_A$ and $Z_V$, as well as their ratio $Z_V/Z_A$ for Shamir DWF. The matching factor $Z_A$ is obtained using the conserved 5d-axial-vector current, whereas the determination of $Z_V$ uses a 3-point function sandwiching a vector current between the same pseudoscalar state on the left and the right. We supplemented these values by our own determinations of $Z_A$ at physical strange quark masses using the 5d-construction for Shamir DWF on some ensembles. 
The extractions of $Z_A$ are shown in Fig.~\ref{Fig.ZAs}. Moreover, we combine our GF-based predictions for the ratio $Z_V/Z_A$ (summarized in Table \ref{Tab.ZchiAZchiV}) with $Z_A$ to deduce the ``missing'' $Z_V$ values on F1S. 

\section{Zero flow time extrapolation}
\label{Sec.ZeroFlowTime}
Taking the $\tau\to 0$ limit is particularly challenging because the data at successive flow times are highly correlated. 
Hence the correlation matrix is highly singular, and calculating its inverse is troublesome. We therefore apply the same procedure as in Ref.~\cite{Black:2025gft}: 
For a chosen range of flow times $(\tau_\text{min},\,\tau_\text{max})$ we perform two uncorrelated fits to all central values first coherently shifted by $+1\sigma$, then by $-1\sigma$. In both fits, the original data uncertainties are used as the weights in the $\chi^2$ function.
This results in an error band having a width greater than or equal to the size of our data point's error bars. To also account for systematic effects due to the chosen flow time range, we independently vary $\tau_\text{min}$ and $\tau_\text{max}$, always performing the two extrapolations described above. 
Specifically, we vary $\tau_\text{min}$ in the range $2/8\lesssim \tau/a^2 \lesssim 6/8$  for fixed $\tau_\text{max}=0.4\,\text{GeV}^{-2}$  as indicated by the vertical dashed and dash-dotted lines in Figs.~\ref{Fig.RatioZchiAZchiV_vs_tau_GeV2} and \ref{Fig.ms}. 
Since in this analysis we first take the $\tau\to 0$ extrapolation for each ensemble, the value of $\tau_\text{min}$ in physical units decreases for finer lattice spacings.
In addition we vary $\tau_\text{max}\in (0.3,\,0.5)\,\text{GeV}^{-2}$ for fixed $\tau_\text{min}/a^2=4/8$, although $\tau_\text{max}$ has only a minor impact on the uncertainty. 
As final error of our extrapolated value we take the full spread of all variations and quote a value with symmetric uncertainty.

\bibliography{draft.bib}
\end{document}